\documentclass[conference]{IEEEtran}
\usepackage[right=0.64in,left=0.63in,top=0.95in,bottom=0.99in]{geometry}
\usepackage{multirow}
\usepackage{booktabs}
\usepackage{amsfonts}
\usepackage{placeins}
\usepackage{array}
\usepackage{amsmath,cases,amssymb}
\usepackage{amsmath}
\usepackage{graphicx}
\usepackage{subcaption}
\usepackage{cite}
\usepackage{epsfig}
\usepackage{todonotes}
\usepackage{url}
\usepackage{algorithm}
\usepackage{algorithmic}
\usepackage{epstopdf}
\usepackage{balance}
\usepackage{color}
\usepackage{algorithm}
\DeclareGraphicsExtensions{.eps,.pdf}

\usepackage{scalerel}
\usepackage{multicol}
\usepackage{amssymb}% http://ctan.org/pkg/amssymb
\usepackage{pifont}% http://ctan.org/pkg/pifont

\usepackage{scalerel}

\let\labelindent\relax
\usepackage{enumitem}

\allowdisplaybreaks

\newcommand{\bseq}{\begin{subequations}}
\newcommand{\eseq}{\end{subequations}}
\newcommand{\baln}{\begin{align}}
\newcommand{\ealn}{\end{align}}
\newcommand{\balnd}{\begin{aligned}}
\newcommand{\ealnd}{\end{aligned}}
\newcommand{\beq}{\begin{equation}}
\newcommand{\eeq}{\end{equation}}
\newcommand{\beqn}{\begin{eqnarray}}
\newcommand{\eeqn}{\end{eqnarray}}
\newcommand{\beqno}{\begin{eqnarray*}}
\newcommand{\eeqno}{\end{eqnarray*}}
\newcommand{\bma}{\begin{displaymath}}
\newcommand{\ema}{\end{displaymath}}
\newcommand{\bnu}{\begin{enumerate}}
\newcommand{\enu}{\end{enumerate}}
\newcommand{\bce}{\begin{center}}
\newcommand{\ece}{\end{center}}
\newcommand{\btb}{\begin{tabular}}
\newcommand{\etb}{\end{tabular}}
\newcommand{\bIEEEeq}{\begin{IEEEeqnarray}}
\newcommand{\eIEEEeq}{\end{IEEEeqnarray}}

\usepackage{scalerel}

\usepackage{caption}

\makeatletter
\newcommand{\linebreakand}{%
\end{@IEEEauthorhalign}
\hfill\mbox{}\par
\mbox{}\hfill\begin{@IEEEauthorhalign}
}
\makeatother

\begin{document}
\title{An Experimental Study of C-Band Channel Model in Integrated LEO Satellite and Terrestrial Systems}
\vspace{-3mm}

\author{\IEEEauthorblockN{Hung Nguyen-Kha${}^{\dagger}$, Vu Nguyen Ha${}^{\dagger}$, Eva Lagunas${}^{\dagger}$, Symeon Chatzinotas${}^{\dagger}$, and Joel Grotz${}^{\ddagger}$}

\IEEEauthorblockA{\textit{${}^{\dagger}$Interdisciplinary Centre for Security, Reliability and Trust (SnT), University of Luxembourg, Luxembourg} \\
       \textit{${}^{\ddagger}$SES S.A., Luxembourg}}
       
       \vspace{-5mm}
       }

\maketitle

\begin{abstract} 
	This paper studies the channel model for the integrated satellite-terrestrial networks operating at C-band under deployment in dense urban and rural areas. Particularly, the interference channel from the low-earth-orbit (LEO) satellite to the dense urban area is analyzed carefully under the impact of the environment's characteristics, i.e., the building density, building height, and the elevation angle. Subsequently, the experimental results show the strong relationships between these characteristics and the channel gain loss. Especially, the functions of channel gain loss are obtained by utilizing the model-fitting approach that can be used as the basis for studying future works of integration of satellite and terrestrial networks (ISTNs).  
\end{abstract}
\vspace{-.2cm}
%\begin{IEEEkeywords}
%\vspace{-.3cm}
%LEO Constellation, Integrated Satellite-Terrestrial Networks, Resource Allocation, User Association. 
%\end{IEEEkeywords}

%\vspace{-1cm}

%\newpage
\section{Introduction} \label{sec:intro}
\vspace{-.1cm}
In recent years, one has been witnessing the rapid growth of traffic demand within terrestrial networks (TNs) due to the blossoming development of internet-based applications. 
Therefore, next-generation mobile networks are expected to satisfy diverse service requirements, including high data rates, massive connectivity, and ubiquitous coverage \cite{SurveyTut_Roadto6G, ProIEEE_6GVision_Challenge_Opp}. However, expanding TNs, especially for reaching underserved and remote areas, is facing several low-cost efficiency difficulties \cite{ISTN_Toward_6G_App_challenge,fontanesi2023artificial,VuHa_ICC23,Juan_PIMRC23,VuHa_PIMRC24}. To manage this surge, the ISTNs emerge as a pivotal solution for achieving global coverage and boosting network capacity \cite{SurveyTut_Roadto6G, ProIEEE_6GVision_Challenge_Opp, Zaid_PIMRC23, Hung_WSA23, hung_ICCW_2023twotier, Hung_2tierLEO}. 
On the other hand, spectrum scarcity poses a significant challenge to expanding capacity, leading both satellites and TNs to explore the same frequency bands, such as the C-band utilized by fixed-satellite-service (FSS) systems. Consequently, this context opens an essential technical issue of investigating cross-system interference.

\subsubsection*{\textbf{Industrial Trend}}
In the recent 3GPP releases, the non-terrestrial-networks (NTNs) have been pressed as an important complementary part of the 5G era to expand the network coverage and provide service continuity \cite{3gpp.38.811}. 
Regarding several NTN-involving 5G use cases, 5G-ALLSTAR project presented in \cite{WCNCW_5G_ALLSTAR_project} focused on some efficient multi-connectivity mechanisms in ISTNs. Subsequently, the project discussed the multi-connectivity and service continuity between NTNs and TNs, considering architecture aspects and system-level assumptions within the 3GPP standards in \cite{3gpp.38.821}. Particularly, coexistence scenarios between 5G New-Radio (NR) TN and NTN are examined in \cite{3gpp.38.863}, highlighting situations where satellites provide direct connectivity to user equipment (UEs) in both rural and urban macro scenarios. Conversely, in dense urban scenarios, NTN UEs are shown to connect to TNs rather than NTNs. The impact of system performance loss due to adjacent band interference is also explored for these coexistence scenarios in the S-band. 
Furthermore, the United States (US) Federal Communications Commission (FCC) has recently proposed measures to support the integration of satellite and terrestrial networks, as detailed in \cite{FCC2322_Supplemental_Space_Coverage}. This proposal emphasizes enhancing TN coverage through supplemental space-based coverage.
Additionally, during the 3GPP TSG RAN R19 workshop, MediaTek Inc. \cite{3gpp.Tdoc_RWS230110} presented insights on opening the TN spectrum bands for SATCOM within the ISTNs.

Following this trend, the utilization of the C-band for the 5G systems including TNs and NTNs has garnered significant interest recently. 
In 2020, the US FCC announced a plan to repurpose a substantial portion of the C-band, previously exclusive to the FSS, for 5G networks in the US. Notably, satellite operators are required to vacate $280$ MHz of the C-band for TN broadband services in December 2023 \cite{FCC20_Cband_5G}.  According
to this transition plan, some of the satellite companies, e.g., SES and Intelsat, have completed the clearing plan in the US before the end of 2023 \cite{SES_Cband_plan, Intelsat_Cband_clear}.
In Europe, the designated bands for 5G include $700$ MHz, $3.6$ GHz, and $24$ GHz, with the $3.6$ GHz band (spanning $400$ MHz from $3.4-3.8$ GHz) serving as the primary frequency for 5G deployment \cite{RSPG23_5Gband}. Consequently, 3GPP has specified sub-bands \textit{n77}, \textit{n78}, and \textit{n79} within the C-band for 5G NR. However, the adoption of the C-band for 5G raises concerns over potential interference with FSS services. Moreover, under the discussed
evolution of ISTNs in \cite{3gpp.38.863}, the TN-NTN coexisting systems could collaboratively utilize the same frequency band, such as C-band, which leads to the need to examine cross-system interference closely. 

\subsubsection*{\textbf{Related Works}}
Recently, several works in literature have studied the cross-system interference issues in the TN-NTN coexisting systems \cite{Eva_Access20, Jumaily_TVT22, Lee_ICSCN21, Lee_TVT23}. 
In \cite{Eva_Access20}, the authors analyzed the interference power caused by 5G base stations (BSs) to the ground FSS receiver in co-channel and adjacent band scenarios. 
Subsequently, the study focuses on the percentage of critical BSs that generate interference power exceeding a predefined threshold.
In \cite{Jumaily_TVT22}, an out-site measurement is carried out to measure the cross-system interference strength from a BS to an exclusive zone near the BS (at a distance of $110$ m). Furthermore, various scenarios involving tuner and filter deployment have been explored for interference analysis. In \cite{Lee_ICSCN21} and \cite{Lee_TVT23}, the authors delved into an NTN-TN spectrum-sharing scenario, with a particular focus on interference from NTN uplink (UL) transmissions. Specifically, \cite{Lee_ICSCN21} examines a reverse spectrum allocation strategy to mitigate aggregated interference from TNs at satellite receivers. Conversely, \cite{Lee_TVT23} introduces a resource group mechanism within the proposed system architecture to manage interference in the NTN UL system, emanating from both TN BSs and inter-beam UEs. However, the analyses in these studies rely on statistical channel models, neglecting the intricate characteristics of dense urban settings.
Moreover, within ISTNs, the satellite's extensive coverage introduces heterogeneity challenges, particularly due to the vast differences in TN scenarios (remote, rural, urban, and dense urban). Consequently, urban architectural features must be accounted for when developing the satellite-to-ground UE (S2G) channel model to accurately reflect the urban environment characteristics.
Subsequently, regarding the interference between earth stations, a new propagation model was proposed in the recommendation ITU-R-452 \cite{ITU_earth_station_interference}. Notably, this model incorporates clutter loss from obstacles like buildings across various environments to enhance accuracy.
However, this statistical model does not account for the specific geographical features of complex environments, such as dense urban areas. Thus, a deeper investigation into the actual environmental characteristics within the channel model is warranted. 

% \vspace{-1mm}
\subsubsection*{\textbf{Motivation and Contribution}}
To the best of our knowledge, the existing studies primarily focus on the detrimental interference from 5G systems to the FSS earth stations. Yet, the analysis of interference to the 5G system caused by FSS downlink transmissions has not received significant attention. This oversight has motivated us to bridge this research gap.
%\subsection{Contributions}
% Regarding the coexisted TN and satellite systems operating in the C-band, the related works aim to analyze the harmful interference caused by the 5G system to the earth stations of the existing FSS system. However, the analysis of interference to the 5G system caused by the FSS downlink transmission has not received significant attention. Furthermore, the channel gain used for the analysis is calculated based on the given model which mainly depends on the relative distance and angle between network components. 
% Regarding the practical scenario, for the low elevation angle at the ground UE located in the dense urban area toward the satellite, especially the LEO satellite, the direct link between the UE and the LEO satellite may not exist due the the blockage, e.g., high building, and only the non-line-of-sight (NLoS) link exists.
% It can be observed, for the dense urban scenario, the channel gain between the UE and the LEO satellite is influenced by the elevation angle at the UE, building height, building density, and beam pattern gain of the LEO satellite toward the UE direction.
The objective of this study is to examine the impact of dense urban area characteristics on the channel model within a realistic C-band coexistence scenario, assessing the viability of utilizing the C-band for ISTNs. Beyond employing a static channel model, we evaluate the channel gain loss attributed to the environment's complex characteristics using a ray tracing tool in MATLAB and actual 3D maps. Our experimental findings reveal significant correlations between these environmental characteristics and channel gain loss. Notably, we derive functions of channel gain loss through a model-fitting approach, providing a foundational framework for future ISTN research. 

%The rest of the paper is organized as follows. The system model is described in Section~\ref{sec:sys_model}. The simulation setup and experimental results are shown in Section~\ref{sec:results}. Finally, the conclusion is summarized in Section~\ref{sec:concl}
%\vspace{-2mm}

\section{System Model} \label{sec:sys_model}
\begin{figure}
	\centering
\includegraphics[width=70mm,height=37mm]{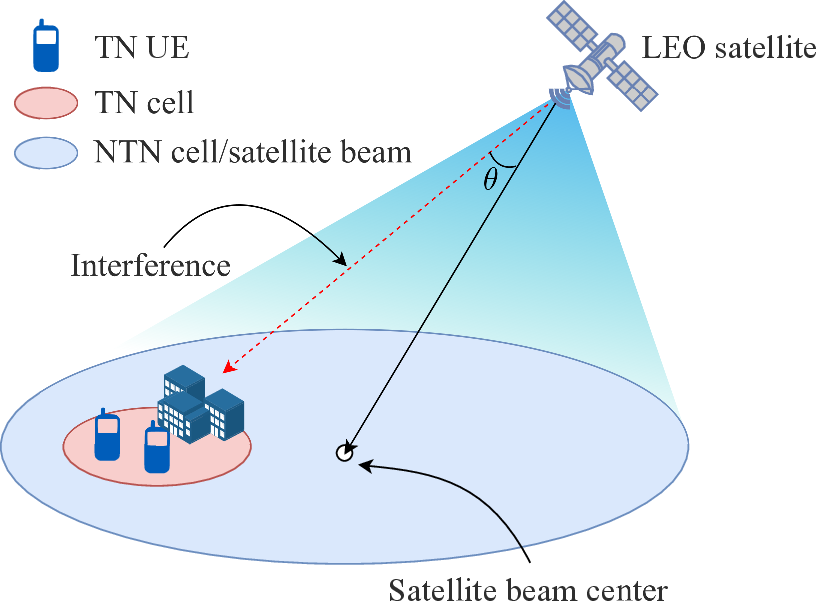}
\captionsetup{font=footnotesize}
	\caption{System Model.}
	\label{fig:system_model}
 \vspace{-3mm}
\end{figure}

\begin{figure*}[!t]
	\centering
	\includegraphics[width=15cm, height=8cm]{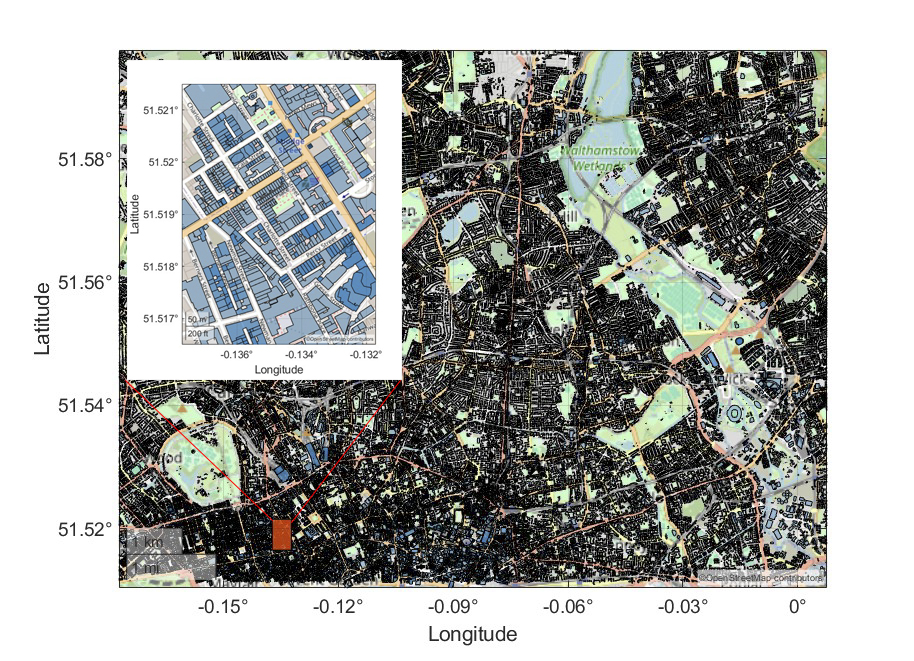}
    \vspace{-3mm}
    \captionsetup{font=footnotesize}
	\caption{Geographic map of an area in London.}
	\label{fig:london_geomap}
    \vspace{-5mm}
\end{figure*}

This work studies the downlink transmission of the coexisting TN and NTN system as Fig.~\ref{fig:system_model}. Similar to the coexisting system described in \cite{3gpp.38.863} (Table 6.1-1), an LEO satellite in NTN is assumed to provide its communication services to the UEs in rural and edge-urban areas while the TN services are only for urban areas. 
In addition, this work focuses on the C-band which both TNs and NTN can access.
Consequently, the potential cross-interference between two networks emerges as a critical issue requiring a thorough examination.
% In particular, due to the large coverage of the satellite formed by the $3-$dB satellite beam radius, the interference power, which is radiated from the satellite toward the urban area direction, can be strong and have a significant effect on TNs.
Particularly, the large coverage of NTN can substantially impact TNs due to the power radiating from the LEO toward urban areas.

% In particular, one LEO satellite serves the NTN cell which is formed by the 3-dB satellite beam coverage. Due to the large coverage of the NTN cell, the TN cell can located inside the NTN cell. One assumes that TN and NTN systems operate in the same frequency band, i.e., C-band, hence, the LEO satellite can cause harmful interference to the TN cell. Furthermore, we assume that the TN cell is deployed in urban areas with high buildings, thus the interference on the TN cell may depend on the area properties.

% Additionally, affected by the high building and high housing density, the channel gain from the satellite to an urban device is different from that of the user located in rural and edge-urban areas. Therefore, this work aims to investigate the S2G links corresponding to the urban area by considering the characteristics of the realistic 3D map, e.g., building height, building density, and elevation angle at the ground receiver points. 
Furthermore, influenced by the presence of tall buildings and high housing density, the channel gain from the LEO to urban devices is different compared to that in rural and edge-urban areas. Therefore, this study aims to analyze the S2G links specific to urban regions, considering various characteristics of a realistic 3D map, such as building height, density, and elevation angle at ground receiver points.
To do so, we first express the overall path loss of an S2G link as
\vspace{-2mm}
\beq
    {PL} = {L}_{\sf{FS}} + {L}_{\sf{r}} + {L}_{\sf{c}} + {L}_{\sf{scen}},
    \vspace{-2mm}
\eeq
wherein ${L}_{\sf{FS}}$ is the free space path-loss which is defined as
\vspace{-2mm}
\beq
    {{L}}_{\sf{FS}} = 20 \log_{10}{f_c} + 20 \log_{10}{d} + 32.45,
    \vspace{-2mm}
\eeq
in which $f_c$ is the operation frequency in GHz, $d$ is the distance between the LEO satellite and the receiver. ${L}_{\sf{r}}$ and ${L}_{\sf{r}}$ are the attenuation due to rain and cloud, respectively. ${L}_{\sf{scen}}$ is the loss due to the properties of scenarios, e.g., dense urban, urban, and rural areas. It is worth noting that the loss ${L}_{\sf{scen}}$ consisting of attenuation due to diffraction, reflection, and wall penetration, is only calculated if the LoS link between the satellite and receiver point is infeasible, which can be calculated as
\vspace{-2mm}
\beq
    L_{\sf{scen}} = \left( {1}/{L_{\sf{ray}}} + {1}/{L_{\sf{wall}}} \right)^{-1},
\eeq
wherein $L_{\sf{ray}}$ is the loss due to diffraction and reflection obtained by ray tracing, and $L_{\sf{wall}}$ is the wall-loss.
It is worth noting the receiver point is placed outside the building, thus the wall penetration loss is modeled twice of O2I building penetration loss which can be calculated in dB scale as \cite{3gpp.38.901}
\vspace{-2mm}
\beq
    {{L}}_{\sf{wall}}= 2 (5 - 10 \log(0.7 \cdot 10^{\frac{-L_{\sf{IRRglass}}}{10}} + 0.3 \cdot 10^{\frac{-L_{\sf{concrete}}}{10}})),
\eeq
where $L_{\sf{IRRglass}}$ and $L_{\sf{concrete}}$ are the penetration loss of IRR glass and concrete materials which are the functions of frequency. It is worth noting that the indoor loss is ignored due to the negligible indoor distance and only the wall-loss is considered.
% Regarding the effect of the obstacle in a real environment (e.g., high building), this work evaluates the resulting loss including reflection and diffraction losses by \textbf{using ray tracing simulation in MATLAB}. The aggregated path loss of a S2G link can be calculated as
Regarding the impact of obstacles in a real environment (e.g., high buildings), this study assesses the resultant loss including reflection and diffraction losses by \textbf{using ray tracing simulation in MATLAB}. The aggregated loss of a S2G link can be computed as
\beq \label{eq: equi_ray_loss}
    % L_{\sf{ray}} = 1/\left( \left| \scaleobj{.8}{\sum_{i=1}^{N}} {e^{-j \phi_{i}}} /{L_{i}} \right | \right),
    L_{\sf{ray}}[\text{dB}] = 10 \log \left( \left| \scaleobj{.8}{\sum_{i=1}^{N}} {e^{-j \phi_{i}}} /{L_{i}} \right |^{-1} \right) - L_{\sf{FS}}[\text{dB}],
    \vspace{-2mm}
\eeq
where $N$ is the ray number simulated by the ray tracing tool, $\phi_{i}$ and $L_{i}$ are the phase shift and propagation loss of $i-$th ray.

% The received interference power can be calculated as
% \beq
%     P_{\sf{I}} = EIRP + 10\log_{10} \psi(\theta) - PL,
% \eeq
% where $\psi(\theta)$ stands for the beam pattern of the LEO satellite antenna, and $\theta$ is the angle between boresight and pointing directions at the viewpoint of the satellite. The beam pattern $\psi(\theta)$ is modeled as
% \beq
%     \psi(\theta) = 
%     \begin{cases}
%         1, & \text{if }\theta=0, \\
%         4 \vert \frac{J_1 ka \sin{\theta}}{ka \sin{\theta}} \vert^2, & \text{otherwise,}
%     \end{cases}
% \eeq
% where $J_1(\cdot)$ is the first-order Bessel function, $a$ is the antenna aperture, $k = 2 \pi f_c / c$.

% \begin{figure*}[h]
% 	\centering
% 	\includegraphics[width=16cm, height=10cm]{Figures/london_geomap_agg.jpg}
% 	\caption{Geographic map of an area in London.}
% 	\label{fig:london_geomap}
% \end{figure*}
%\vspace{-2mm}
\section{Experimental Results} \label{sec:results}
This section introduces the simulation for an area in London city and provides the experimental results of the path loss between an LEO satellite and the receiver points. The following is to describe the simulation setup and experimental results.

\begin{figure}[!t]
	\centering
	\includegraphics[width=9cm, height=4.5cm]{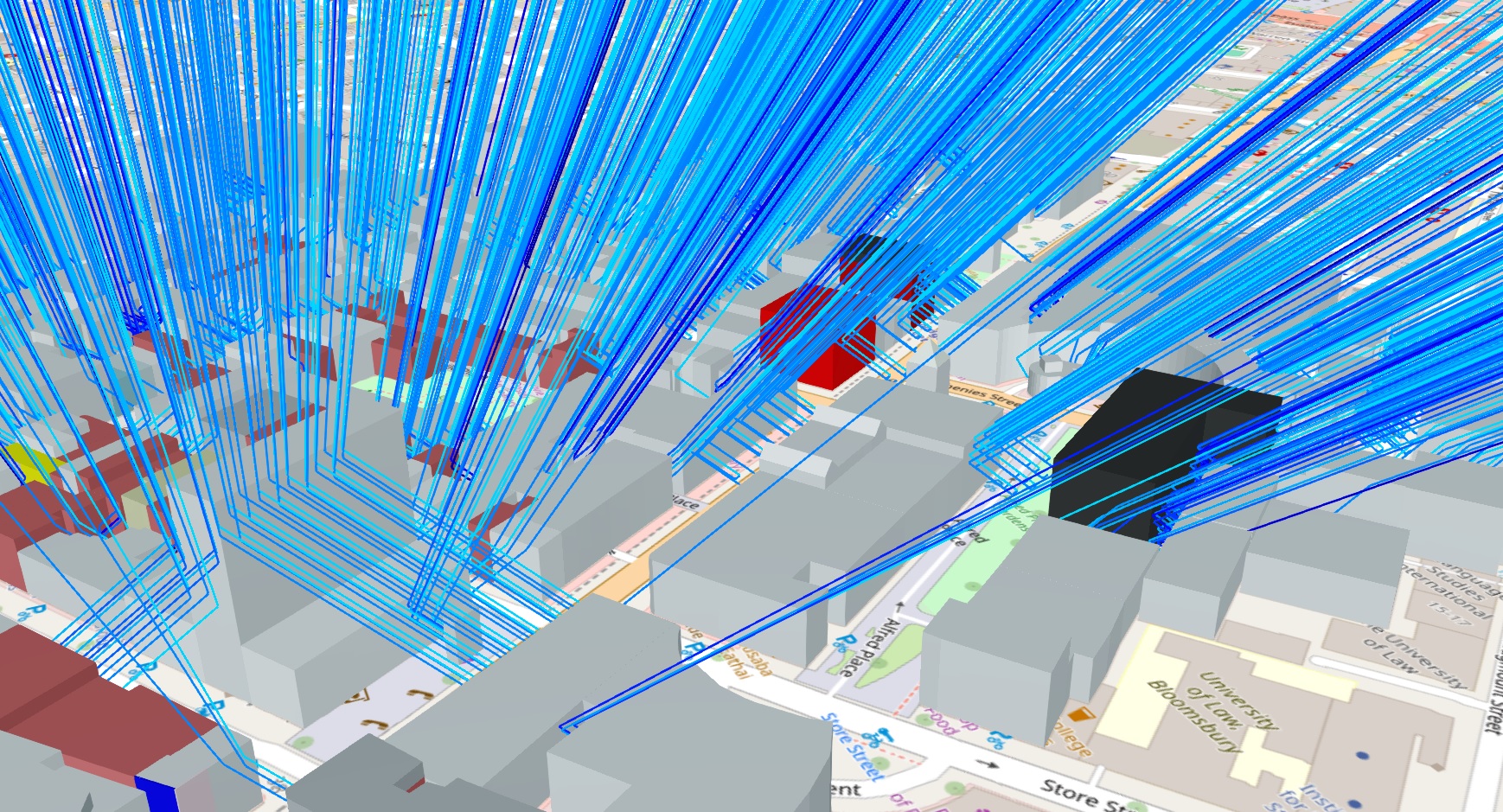}
 \captionsetup{font=footnotesize}
	\caption{Ray trace results for a 3D map region in London.}
	\label{fig:ray_3d}
    \vspace{-3mm}
\end{figure}

\begin{table}[!t]
\captionsetup{font=footnotesize}
	\caption{Simulation Parameters}
	\label{tab:parameter}
	\centering
	%\setlength{\tabcolsep}{0.695em}
	%\setlength{\extrarowheight}{0.6em}
	%	\scalebox{0.8}{
	\begin{tabular}{l|l}
		\hline
		Parameter & Value \\
		\hline\hline
        Considered latitude limitation              & $[51.5115 ^\circ \text{N}, 51.5965^\circ \text{N}]$   \\
        Considered longitude limitation              & $[0.1772^\circ \text{W} , 0.0076^\circ \text{E}]$   \\
        Size of a map segment                   & $0.005^\circ \times 0.005^\circ$ \\
        LEO satellite altitude                  & $550$ km \\
        Receiver height                         & $1$ m   \\
        Size of receiver point mesh in one segment & $100 \times 100$ points      \\
        Operation frequency, $f$              & $3.4$ GHz \\
        Penetration loss of IRR glass, $L_{\sf{IRRglass}}$ & $23 + 0.3f$[GHz]    \\
        Penetration loss of concrete, $L_{\sf{concrete}}$ & $5 + 4f$[GHz]    \\
		\hline		   				
	\end{tabular}
	%	\vspace{-0.5pt}
\end{table}

\begin{figure*}[!t]
	\centering
	\includegraphics[width=15cm, height=8cm]{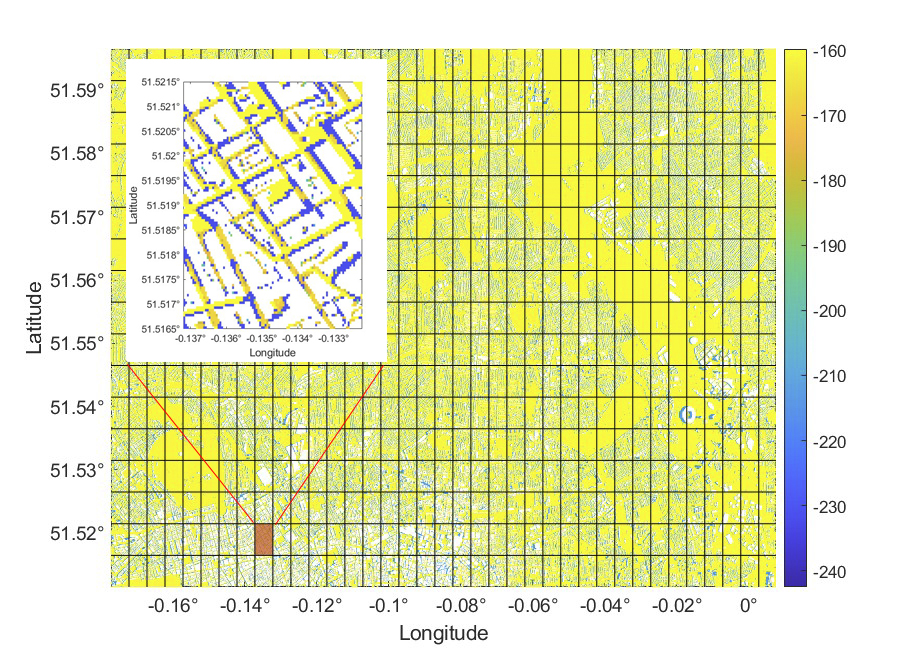}
    \vspace{-3mm}
    \captionsetup{font=footnotesize}
	\caption{Heat map of path-loss in an area in London [dB].}
	\label{fig:london_heatmap}
    \vspace{-6mm}
\end{figure*}
\subsection{Simulation Setup}
The simulation is conducted within an area in London bounded by latitude and longitude coordinates ranging $[51.5115 ^\circ \text{N}, 51.5965^\circ \text{N}]$ and $[0.1772^\circ \text{W} , 0.0076^\circ \text{E}]$, respectively. The geographic map of this area is illustrated in Fig.~\ref{fig:london_geomap}. 
However, due to the heterogeneity of the vast expanse, the area is partitioned into a grid of segments, wherein the latitude and longitude sizes of each segment are $0.005^\circ$.  For instance, a detailed geographic map of a segment located at row $2$, column $9$ of the map grid is depicted in Fig.~\ref{fig:london_geomap}. 
Furthermore, in order to enhance the accuracy of path loss assessment, a mesh of receiver points is generated for each map segment.
Subsequently, the average path loss between the satellite and a map segment is computed by averaging the results obtained from its receiver points.
The rain attenuation $L_{\sf{r}}$ and the cloud attenuation $L_{\sf{c}}$ are set as specified in \cite{rain_attenuation} and \cite{cloud_attenuation}, respectively.
The key simulation parameters are described in Table.~\ref{tab:parameter}. For brevity, let us denote the average building height as $h$, the building density as $\mu$, and the elevation angle as $\theta_{\sf{elev}}$.

\subsection{Simulation Results}
As mentioned above, the evaluation is carried out for each map segment due to the heterogeneity of the large area. Fig.~\ref{fig:ray_3d} shows the 3D map and the ray tracing result of the map segment at row $2$ column $9$ whose geographic map is shown in Fig.~\ref{fig:london_geomap}. Subsequently, the path loss result for each receiver point is determined based on the obtained ray tracing results and \eqref{eq: equi_ray_loss}.
\begin{figure}[!t]
	\centering
 \captionsetup{font=footnotesize}
    \begin{subfigure}{0.5\textwidth}
        \centering
        \includegraphics[width=80mm,height=50mm]{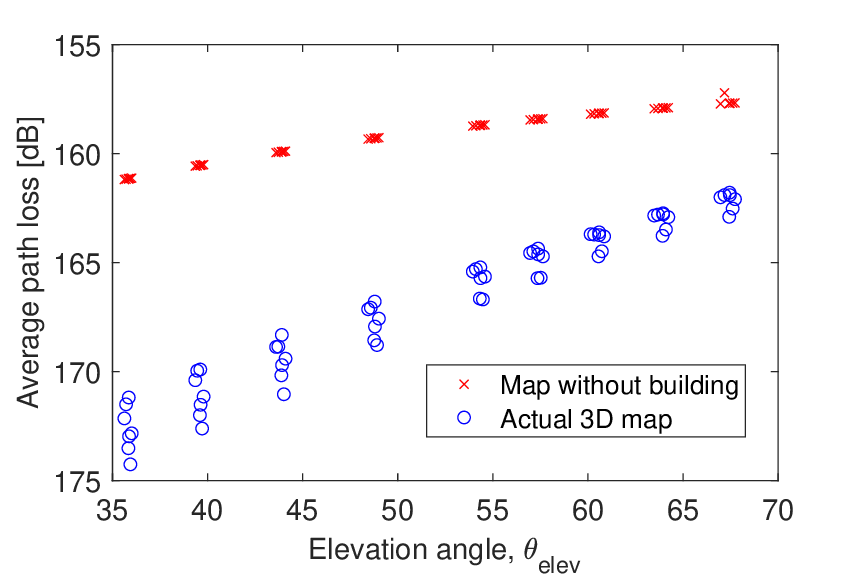}
        %\vspace{-6mm}
        \captionsetup{font=footnotesize}
        \vspace{-1mm}
        \caption{Path-loss.}
        \label{fig:PL_elev}
        % \vspace{-1mm}
    \end{subfigure}
    % \vspace{-2mm}
    % \hfill
    \begin{subfigure}{0.5\textwidth}
        \centering
        \includegraphics[width=80mm,height=50mm]{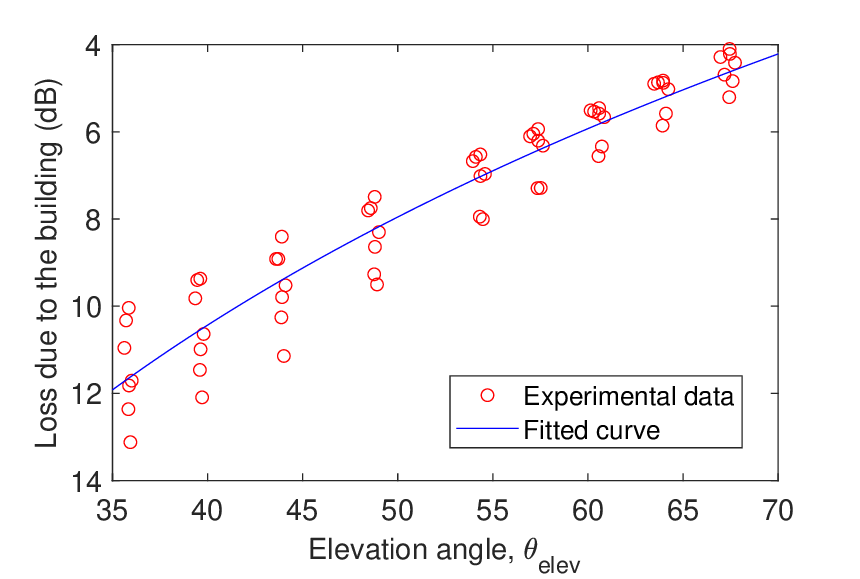}
        %\vspace{-6mm}
        \captionsetup{font=footnotesize}
        \vspace{-1mm}
        \caption{Channel gain loss.}
        \label{fig:Loss_elev}
        \vspace{-1mm}
    \end{subfigure}
    \vspace{-4mm}
\captionsetup{font=footnotesize}
	\caption{The path-loss versus the elevation angle with the building density $\mu = 0.3$ and the average building height $h = 8.9$ m.}
     \label{fig:result_elev}
 \vspace{-2mm}
\end{figure}

Fig.~\ref{fig:london_heatmap} depicts the heat map of the path loss result of the whole considered area and the map segment corresponding to Fig.~\ref{fig:london_geomap} at the elevation angle is $40^\circ$. In the heat map, white areas represent buildings, while other colors indicate varying degrees of path loss. Notably, there's a significant disparity in path loss results between segments characterized by high building density and those with minimal buildings.
In segments with few or no buildings, the channel gain between the satellite and these segments appears strong and relatively uniform over the whole segment, the path loss is about $-160$ dB. Conversely, in the segments with high and dense buildings, the channel gains from the satellite to the receiver points within each segment are heterogeneous due to the effect of the building. 
For instance, let's consider the heat map result of one segment at row $2$ column $9$, receiver points with LoS links to the satellite demonstrate path loss results similar to segments without buildings, around $-160$ dB. 
However, at receiver points lacking LoS links with the satellite, the channel gain is considerably lower, i.e., about from $-170$ dB to $-180$ dB of path loss, or the channel is nearly blocked, i.e., about from $-200$ dB to $-240$ dB of path loss. \vspace{-1mm}

To gain further insights into the relationship between the environmental characteristics and path loss results, we analyze the effects of building height, building density, and elevation angle in the following parts. Specifically, the simulation is conducted under two scenarios: "Map without building" and "Actual 3D map".
The impact of the elevation angle on the path loss results is shown as per Fig.~\ref{fig:result_elev}, focusing on segments with building density $\mu = 0.3$ and the average building height of $h = 8.9$ m. First, the average path-loss result of the two examined cases, i.e., the map without building and the actual 3D map, is shown as per Fig.~\ref{fig:PL_elev}. It can be foreseen that the higher elevation angle results in stronger channel gain in both cases due to the closer slant range between the satellite and the receiver points. On the other hand, due to the effect of the building when considering the actual 3D map in the latter case, the channel gain in the first case is stronger than that of the other at all examined elevation angle values. Especially, the gap of loss between the two cases is larger as the lower elevation angle, 
since the lower elevation angle leads to a lower probability of LoS link due to the building obstruction which can be observed as per Fig.~\ref{fig:ray_3d}.
% which indicates the stronger impact of the building on the channel gain. 
% This can be explained intuitively by the ray tracing results in Fig.~\ref{fig:ray_3d} as follows. For a given 3D map area, the lower elevation angle leads to a lower probability of LoS link due to the obstruction of the buildings. Hence, the average channel gain over the whole area is reduced.
Subsequently, the channel gain loss result is shown as per Fig.~\ref{fig:Loss_elev}. 
The loss due to the building obstruction seems to decrease logarithmically as the elevation angle increases. 
The loss of channel gain is significant at the low elevation angle. Particularly, at the elevation angle lower than $45^\circ$, the loss value is about $10$~dB or above. In addition, it can be seen that the loss value is more dispersal at the lower elevation angle due to the stronger impact of the building and the complexity of the environment. Hence, we utilize the curve-fitting model in Matlab \cite{VuHa_TWC21,VuHa_TWC20,VuHaGC2022} to approximate the mathematical relationship between the loss due to buildings and the elevation angle.
In particular, as the observed logarithmic trend, one can assume that the loss follows a logarithmic function of the form as 
\beq \label{eq: loss elev}
    % PL_{\sf{scen,elev}} \text{[dB]} = 24.44 \log(\theta_{\sf{elev}}) - 49.04.
    PL_{\sf{scen,elev}} \text{[dB]} = a_{1}^{\sf{elev}} \log(\theta_{\sf{elev}}) + a_{2}^{\sf{elev}}.
\eeq
The fitted curve of average loss versus the elevation angle is depicted as Fig.~\ref{fig:Loss_elev} wherein the coefficients are obtained as $a_{1}^{\sf{elev}} = -25.6$ and $a_{2}^{\sf{elev}} = 51.44$.

\begin{figure} [h]
	\centering
    \begin{subfigure}{0.5\textwidth}
        \centering
        \includegraphics[width=80mm,height=50mm]{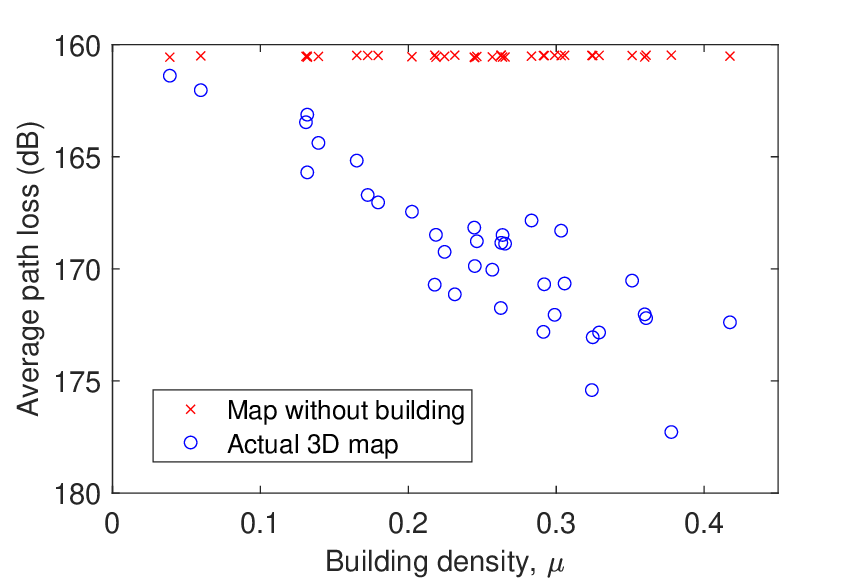}
        %\vspace{-6mm}
        \captionsetup{font=footnotesize}
        \vspace{-1mm}
        \caption{Path-loss.}
        \label{fig:PL_dense}
        \vspace{-1mm}
    \end{subfigure}
    % %\vspace{2mm}
    % \hfill
    \begin{subfigure}{0.5\textwidth}
        \centering
        \includegraphics[width=80mm,height=50mm]{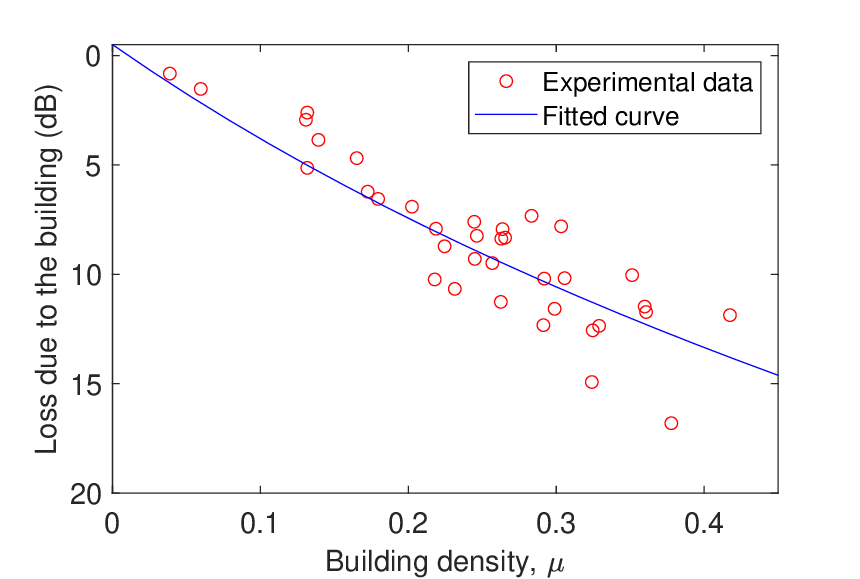}
        %\vspace{-6mm}
        \captionsetup{font=footnotesize}
        \vspace{-1mm}
        \caption{Channel gain loss.}
        \label{fig:Loss_dense}
        \vspace{-1mm}
    \end{subfigure}
    \vspace{-4mm}
    \captionsetup{font=footnotesize}
	\caption{The path-loss versus the building density with the average building height of $h = 8.9$ m and elevation angle $\theta_{\sf{elev}} = 40^{\circ}$.}
    \label{fig:result_dense}
\vspace{-2mm}
\end{figure}

Subsequently, we investigate the impact of the building density on the path loss as per Fig.~\ref{fig:result_dense} for the average building height of $h = 8.9$~m and the elevation angle of $\theta_{\sf{elev}} = 40^\circ$. First, the average path loss results in two considered cases are presented in Fig.~\ref{fig:PL_dense}. It can be seen that the channel gain in the first case is nearly constant across the examined segments since it only depends on the terrain height. In contrast, the path loss varies significantly over the building density of the segment in the 3D map case. In particular, as the building density increases, the gap in terms of path loss between the two cases widens quickly in the range of building density from about $0$ to $0.25$ whereas it increases at a slower rate in higher building densities. 
At the building density of about $0.05$, the gap in terms of the path-loss between two considered cases is negligible, i.e., only about $1$~dB, since the minimal difference of map between the two cases. However, the effect of the building on the path loss is clearer when the building is more dense. 
For more insight into the impact of building density, the loss result versus building density is depicted as per Fig.~\ref{fig:Loss_dense}. 
It can be seen that a sufficient value of building density causes a significant loss to the channel gain. For instance, there are the segments that experience the loss due to the building of about $10$~dB and $17$~dB corresponding to the building density of $0.24$ and $0.37$, respectively. To further elucidate the mathematical relationship between the building density and the loss, the fitting approach is applied to the experimental data, wherein the relationship function is assumed to be a logarithmic form as
\beq
    PL_{\sf{scen,den}} = a_{1}^{\sf{den}} \log(\mu + a_{2}^{\sf{den}}) + a_{3}^{\sf{den}}.
    \vspace{-1mm}
\eeq
The obtained curve is illustrated as per Fig.~\ref{fig:Loss_dense} with $a_{1}^{\sf{den}} = 53.76, a_{2}^{\sf{den}} = 0.49$ and $a_{3}^{\sf{den}} = 15.96$.

% \begin{figure}
% 	\centering
% 	\includegraphics[width=8cm]{Figures/PL_dense.eps}
% 	\caption{The average path-loss versus the building density with the average building height of $8.9$ m and elevation angle $40^{\circ}$.}
% 	\label{fig:PL_dense}
% \end{figure}

% \begin{figure}
% 	\centering
% 	\includegraphics[width=8cm]{Figures/Loss_dense.eps}
% 	\caption{The average loss versus the building density and the fitted curve with the average building height of $8.9$ m and elevation angle $40^{\circ}$.}
% 	\label{fig:Loss_dense}
% \end{figure}

\begin{figure} 
	\centering
    \begin{subfigure}{0.5\textwidth}
        \centering        \includegraphics[width=80mm,height=50mm]{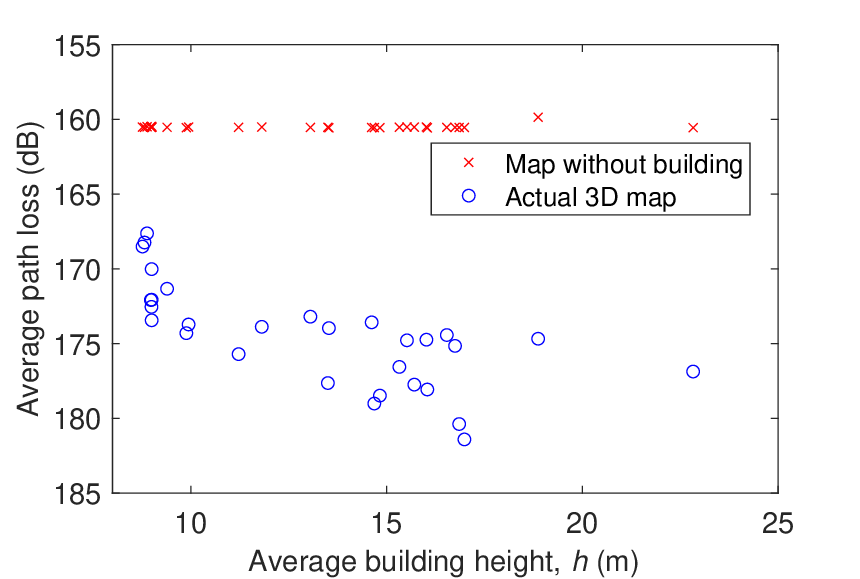}
        %\vspace{-6mm}
        \captionsetup{font=footnotesize}
        \vspace{-1mm}
        \caption{Path-loss.}
        \label{fig:PL_hbuilding}
        % %\vspace{-2mm}
    \end{subfigure}
    % %\vspace{2mm}
    % \hfill
    \begin{subfigure}{0.5\textwidth}
        \centering
        \includegraphics[width=80mm,height=50mm]{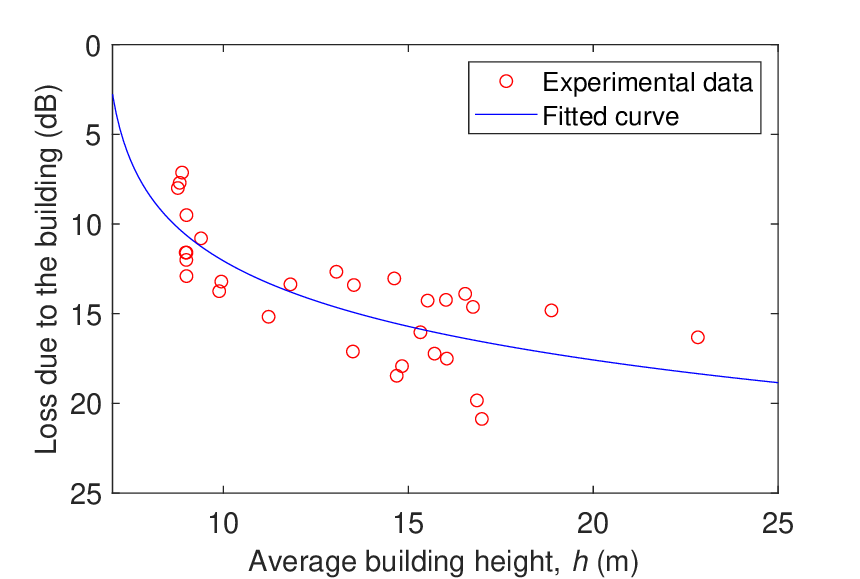}
        %\vspace{-6mm}
        \captionsetup{font=footnotesize}
        \vspace{-1mm}
        \caption{Channel gain loss.}
        \label{fig:Loss_hbuilding}
        \vspace{-1mm}
    \end{subfigure}
    \vspace{-2mm}
\captionsetup{font=footnotesize}
	\caption{The path-loss versus the average building height with the building density of $\mu = 0.5$ and elevation angle of $\theta_{\sf{elev}} = 40^{\circ}$.}
 \label{fig:result_hbuilding}
 %\vspace{-3mm}
\end{figure}

Finally, we investigate the influence of the average building height on the path loss as follows in the scenario of building density $\mu = 0.4$ and elevation angle $\theta_{\sf{elev}} = 40^{\circ}$.  
First, the path loss result in two cases of the map is illustrated as per Fig.~\ref{fig:PL_hbuilding}. Similar to Fig.~\ref{fig:PL_dense}, the path loss results remain nearly unchanged in the case without building. In contrast, the path loss result varies clearly across segments due to the influence of buildings in the case of the actual map. It can be observed intuitively that as the increasing of average building height, the channel gain decreases first quickly and then more slowly. 
% To assess the impact of the building height on the channel gain loss, the average loss versus the average building height corresponding to the result in Fig.~\ref{fig:PL_hbuilding} is shown in Table.~\ref{tab:PL_hbuilding}. It can be seen that the building height affects significantly the channel gain. In particular, the change of the average building height from $8.8$~m to $16.5$~m results in the increase of the average loss from $7.4$~dB to $17.8$~dB.    
To assess the impact of the building height on the channel gain loss, the loss versus the average building height corresponding to the result in Fig.~\ref{fig:PL_hbuilding} is depicted in Fig.~\ref{fig:Loss_hbuilding}. It can be seen that the building height affects significantly the channel gain. In particular, the change of the average building height from $8.8$~m to $13.5$~m results in the increase of the average loss from $8$~dB to $17$~dB. Based on the observed varying trend, one assumes that the experimental data points are approximated as
%\vspace{-2mm}
\beq \label{eq: loss h}
    % PL_{\sf{scen,h}} \text{[dB]}= -9.2 \log(-6.7 + h \text{[m]}) - 7.3,
    PL_{\sf{scen,h}} \text{[dB]}= a_{1}^{\sf{h}} \log(h \text{[m]} + a_{2}^{\sf{h}}) + a_{3}^{\sf{h}}.
\eeq
For the fitting model, the fitted function is illustrated as per Fig.~\ref{fig:Loss_hbuilding} wherein $a_{1}^{\sf{h}} = 9.2, a_{2}^{\sf{h}} = -6.7$ and $a_{3}^{\sf{h}} = 7.3$.

% \begin{figure}[!h]
% 	\centering
% 	\includegraphics[width=8cm]{Figures/PL_hbuilding.eps}
% 	\caption{The average path-loss versus the average building height with the building density of $0.5$ and elevation angle of $40^{\circ}$.}
% 	\label{fig:PL_hbuilding}
% \end{figure}

% \begin{figure}[!h]
% 	\centering
% 	\includegraphics[width=8cm]{Figures/Loss_hbuilding.eps}
% 	\caption{The average loss versus the average building height and the fitted curve with the building density of $0.5$ and elevation angle of $40^{\circ}$.}
% 	\label{fig:Loss_hbuilding}
% \end{figure}

% In addition, it can be seen that the channel gain between the LEO satellite and the receivers located in the urban areas is impacted significantly by the characteristics of the urban environment and the elevation angle at the receiver. Regarding the considered ISTN scenario, the channel gain can be weak corresponding to the specific location which depends on the density and height of buildings, and the elevation angle, which results in the negligible interference from the LEO satellite to the terrestrial receiver.
% Summarily, the obtained approximated functions of loss due to the scenario characteristics in different cases are shown in Table.~\ref{tab:fit func}. It is worth noting that these function forms can be reused for the other experimental data with the other coefficients. Especially, the derived function can be used as the basis for studying the ISTNs.

\begin{table}[!t]
\captionsetup{font=footnotesize}
	\caption{Approximated functions of the loss.}
	\label{tab:fit func}
	\centering
    \scalebox{0.9}{
	\begin{tabular}{|l|l|l|l}
		\hline
		Scenario & Function & Coefficient \\
		\hline\hline
        \shortstack[l]{$\mu = 0.3$ \\ $h = 8.9$ m} & \shortstack[l]{$PL_{\sf{scen,elev}} \text{[dB]} $ \\ $ \quad = a_{1}^{\sf{elev}} \log(\theta_{\sf{elev}}) + a_{2}^{\sf{elev}}$} & \shortstack[l]{ $a_{1}^{\sf{elev}} = -25.6$ \\ $ a_{2}^{\sf{elev}} = 51.44$ } \\
		\hline		   				
        \shortstack[l]{$h = 8.9$ m \\ $\theta_{\sf{elev}} = 40^{\circ}$} & \shortstack[l]{$PL_{\sf{scen,den}} \text{[dB]} $ \\ $ \quad = a_{1}^{\sf{den}} \log(\mu + a_{2}^{\sf{den}}) + a_{3}^{\sf{den}}$} & \shortstack[l]{ $a_{1}^{\sf{den}} = 53.73, a_{2}^{\sf{den}} \!\! = \! 0.49$ \\ $a_{3}^{\sf{den}} = 15.96$} \\
		\hline		   				
        \shortstack[l]{$\mu = 0.4$ \\ $\theta_{\sf{elev}} = 40^{\circ}$} & \shortstack[l]{$PL_{\sf{scen,h}} \text{[dB]} $ \\ $ \quad = a_{1}^{\sf{h}} \log(h \text{[m]} + a_{2}^{\sf{h}}) + a_{3}^{\sf{h}}$} & \shortstack[l]{ $a_{1}^{\sf{h}} = 9.2, a_{2}^{\sf{h}} = -6.7$ \\ $a_{3}^{\sf{h}} = 7.3$} \\
		\hline		   				
	\end{tabular}
 }
	%\vspace{-2mm}
\end{table}

In addition, it can be seen that the channel gain between the LEO satellite and the receivers located in the urban areas is impacted significantly by the characteristics of the urban environment and the elevation angle at the receiver. Regarding the considered ISTN scenario, the channel gain may weaken at specific locations due to factors such as building density and height, as well as the elevation angle, which results in negligible interference from the LEO satellite to the terrestrial receiver.
Summarily, the obtained approximated functions of loss due to the scenario characteristics across various cases are presented in Table.~\ref{tab:fit func}. It is noteworthy that these function forms can be applied to the other experimental data with the other coefficients. Moreover, the derived function can be used as the basis for studying the ISTNs to consider the impact of the characteristics of the actual map.

%\vspace{-2mm}
\section{Conclusion} \label{sec:concl}
% In this work, we studied the integrated LEO satellite and terrestrial downlink systems in heterogeneous areas, rural and dense urban areas. Subsequently, the channel gain between the LEO satellite and the dense urban area is investigated with the consideration of complex characteristics of the urban environment and different scenarios of elevation angle. The experimental results show the significant impact of the urban area properties and the elevation angle on the channel gain which verifies the feasibility of the considered ISTN scenario.
% Furthermore, the obtained function of the channel gain loss can be used as the basis for future works on ISTNs.
In this work, we studied the integrated LEO satellite and terrestrial downlink systems in a heterogeneous region encompassing both rural and densely urban areas. Specifically, we delved into the S@G channel gain for dense urban regions, taking into account the intricate characteristics of dense urban environments and various scenarios of elevation angles. Our experimental findings underscore the substantial influence of urban architectural characteristics and elevation angles on channel gain, thereby affirming the feasibility of the ISTN scenarios. 
Notably, the derived function of the channel gain loss can be used as the basis for future works on ISTNs.

\vspace{-1mm}
\section*{Acknowledgment}
\vspace{-1mm}
This work has been supported by the Luxembourg National Research Fund (FNR) under the project INSTRUCT (IPBG19/14016225/INSTRUCT). 
%and project MegaLEO (C20/IS/14767486).

\vspace{-1mm}
\bibliographystyle{IEEEtran}
%\balance
\bibliography{Journal}
\end{document}